\documentclass[aps,twocolumn]{revtex4}
\usepackage{graphicx}   
\usepackage{latexsym}   

\begin{document}

\newcommand{\note}[1]{\marginpar{\tiny {#1}}}   
\newcommand{\bld}[1]{\mbox{\boldmath $#1$}}     
\newcommand{\etal}{{\em et al.}}			
\newcommand{\ie}{{\em i.e.}}				
\newcommand{\code}{{\sc\small FREYA}}
\newcommand{\beq}{\begin{equation}}
\newcommand{\eeq}{\end{equation}}
\newcommand{\beqar}{\begin{eqnarray}}
\newcommand{\eeqar}{\end{eqnarray}}
\newcommand{\half}{\mbox{{$1\over2$}}}
\newcommand{\SKIP}[1]{{ }}

\title{Coupled fission fragment angular momenta}

\author{J\o rgen Randrup$^1$}

\affiliation{
  $^1$Nuclear Science Division,
  Lawrence Berkeley National Laboratory,
  Berkeley, CA 94720, USA}

\date{August 20, 2022}

\begin{abstract}
Nuclear fission produces fragments whose spins are coupled 
to the relative angular motion via angular momentum conservation.
It is shown how ensembles of such spins can readily be obained
by either direct microcanonical sampling
or by sampling of the associated normal modes of rotation.
The resulting distribution of the spin-spin opening angle
is illustrated in various three- and two-dimensional scenarios
and it is demonstrated how recent mutually conflicting model calculations
can be well reproduced with different assumptions about the scission geometry.
\end{abstract}

\maketitle

\section{Introduction}
\label{intro}

Nuclear fission has become a very active topic,
both experimentally \cite{ANS,SJ2018} and theoretically \cite{SJ2018,FoF}.
In partiucular, there has recently been considerable interest in the 
calculation of the correlations between the angular momenta of the fragments 
and a number of mutually contradictory predictions have been made
about the distribution of the spin-spin opening angle, $P_{12}(\psi)$.

The first calculations of $P_{12}(\psi)$ were made with the fission
event generator \code\
which assumes that the fragment spins are perpendicular to the fission axis.
It was found that, apart from the restriction of being two-dimensional,
the spins were nearly independent, in magnitude as well as direction
\cite{VogtPRC103,RandrupPRL127}.
Accordingly, $P_{12}(\psi)$ exhibited only 
a small undulation away from constancy.

Subsequently, using time-dependent density functional theory
with various energy-density functionals,
Bulgac \etal\ \cite{BulgacPRL128} found that $P_{12}(\psi)$ 
exhibits a large angular variation and peaks around $\psi\approx130^\circ$.

Very recently, dynamical calculations with Antisymmetrized Molecular Dynamics
have yielded a nearly symmetric distribution that peaks 
slightly above $90^\circ$ \cite{AMD}.

The present situation is thus rather unclear
and it is the purpose of this paper to provide a framework
within which it can be understood how such widely different results 
can emerge when the coupled spins are sampled under different assumptions.

We start by describing two (different but equivalent) general techniques
for sampling angular momenta that are subject to conservation relations
that render them correlated.
Though the methods are applicable generally,
we concentrate here on the sampling of the two fragment spins 
$\bld{S}_1$ and $\bld{S}_2$ together with the angular momentum
associated with the relative fragment motion, $\bld{S}_0$.

Then these methods are applied to the sampling of the three angular momenta,
$\{\bld{S}_i\}$,
demonstrating that the two methods do indeed yield identical results.
We first consider the more general scenario
in which the angular momenta are three-dimensional vectors
that are constrained only by the conservation laws.
Subsequently, we address the scenario
in which the angular momenta must also be perpendicular to the fission axis
(as certainly $\bld{S}_0$ must be by definition), a requirement that
effectively reduces the spins to being two-dimensional.

\section{Sampling methods}

Generally, the rotational degrees of freedom of the fledging fragments 
can exchange energy with the remainder of the system.
Because the associated rotational energies are typically relatively small 
in comparison with the internal excitation energy,
the nuclear complex effectively acts as an energy reservoir.
Therefore, in the present study where the energy is not important,
we shall assume that the rotational energies have canonical distributions
characterized by the prevailing temperature in the system at scission.
By contrast, because no external torques are acting on the fissioning system,
its overall angular momentum is conserved.
For simplicity and with no bearing on the results,
it is assumed that the total angular momentum vanishes,
so angular momentum conservation requires 
$\bld{S}_1+\bld{S}_2+\bld{S}_0=\bld{0}$.

We describe below two different but equivalent methods
for sampling the microcanonical ensemble
of the three coupled angular momenta $\{\bld{S}_i\}$.

An important role is played by the moments of inertia
whose relative magnitudes influence the appearance of $P_{12}(\psi)$.
For simplicity we treat the two fragments as solid spheres.
Their moments of inertia are denoted by ${\cal I}_1$ and ${\cal I}_2$ 
and the numerical calculations use ${\cal I}_i=\mbox{$2\over5$}M_iR_i^2$
where $M_i$ is the fragment mass and $R_i$ is its radius.
Furthermore, ${\cal I}_0=\mu R^2$ is
the moment of inertia associated with the relative motion 
where $R=R_1+R_2+d$ is the distance between the two fragment centers
and $\mu=M_1M_2/(M_1+M_2)$ is the reduced mass.

\subsection{Direct microcanonical sampling}
\label{microcan}

The perhaps conceptually simplest method
makes a direct sampling of the microcanonical ensemble
defined by the total energy $E$ and the total angular momentum.

The expectation value of any spin-dependent ``observable'', $F\{\bld{S}_i\}$,
is then given by
\beqar\label{micro} \nonumber
\langle F\{\bld{S}_i\}\rangle &=& {!\over\Omega_D(E)}
\prod_{i=0}^2\left[\int d^D\bld{S}_i\right]\, F\{\bld{S}_i\}\\ &\times&
\delta(E-\sum_{i=0}^2{S_i^2\over2{\cal I}_i})\,\,
\delta^{(D)}(\sum_{i=0}^2\bld{S}_i)\ ,
\eeqar
where the corresponding microcanonical phase-space volume 
for $D$-dimensional spins is \cite{RandrupCPC59}
\beqar \nonumber
\Omega_D(E)\! &\equiv&\!
\prod_{i=0}^2\left[\int d^D\bld{S}_i\right]\,
\delta(E-\sum_{i=0}^2{S_i^2\over2{\cal I}_i})\,
\delta^{(D)}(\sum_{i=0}^2\bld{S}_i)\\
&=&\! {2\pi\over\Gamma(D)}
\left({{\cal I}_1{\cal I}_2{\cal I}_0\over{\cal I}}\right)^{D/2}
\left[2\pi E\right]^{D-1},
\eeqar
with ${\cal I}\equiv{\cal I}_1+{\cal I}_2+{\cal I}_0$.
Although the total rotational energy $E$ fluctuates in the actual fissioning
system, this has no impact when only directional effects are considered.
The specific value of $E$ is thus immaterial.
The focus is here on the opening angle between
the two fragment spins, $\psi_{12}$,
where $\cos\psi_{12}=\bld{S}_1\cdot\bld{S}_2/(S_1S_2)$,
so the normalized opening-angle distribution $P_{12}(\psi)$ is given by 
the expectation value of $F\{\bld{S}_i\}\equiv\delta(\psi_{12}-\psi)$.
 
The actual evaluation is carried out by 
sampling the microcanonical distribution.
Though this may appear to be a technically demanding task,
it can in fact be accomplished remarkably easily 
\cite{RandrupCPC59,RandrupNPA522}:
first tentative spin values $\{\bld{S}_i'\}$ are sampled independently 
from Boltzmann distributions with a common but arbitrary temperature;
then the resulting total angular momentum is calculated,
$\bld{S}'=\bld{S}_1'+\bld{S}_2'+\bld{S}_0'$
and the corresponding rotational frequency is determined,
$\bld{\omega}'=\bld{S}'/{\cal I}$;
the overall rotational motion is then removed,
yielding $\bld{S}_i''=\bld{S}_i'-{\cal I}_i\bld{\omega}'$,
and these spins are finally scaled by a common factor $c=\sqrt{E/E''}$
to ensure that the specified total energy $E$ is matched,
yielding the spins $\{\bld{S}_i\}=\{c\bld{S}_i''\}$.
(This last step is of course superfluous when only directional effects
are of interest.)
The resulting spins clearly satisfy the requirements
on the total angular momentum and, crucially,
they are distributed according to the correct microcanionical measure
(see Refs.\ \cite{RandrupCPC59,RandrupNPA522} for a proof of this key feature).
This sampling method is {\em efficient} (no rejections are required), 
{\em robust} (no delicate numerical cancellations occur), and
{\em fast} (millions of samples can be obtained in seconds on a typical laptop).

\subsection{Sampling of normal modes}
\label{modes}

An alternative, equivalent, sampling technique
utilizes the normal spin modes of the system
which are obtained by bringing the rotational energy onto diagonal form,
\beq\label{E}
E = {S_1^2\over2{\cal I}_1} +{S_2^2\over2{\cal I}_2}
  + {|\bld{S}_1\!+\bld{S}_2|^2\over2{\cal I}_0} 
= {s_+^2\over2{\cal I}_+}
+ {s_-^2\over2{\cal I}_-},
\eeq
where angular momentum conservation has been used to replace
the angular momentum of the relative motion, $\bld{S}_0$,
by $-\bld{S}_1-\bld{S}_2$.
The moments of inertia of the normal modes are \cite{RDV}
\beq
{\cal I}_+^{-1}=[{\cal I}_1+{\cal I}_2]^{-1}+{\cal I}_0^{-1}\ ,\,\
{\cal I}_-^{-1}={\cal I}_1^{-1}+{\cal I}_2^{-1}\ ,
\eeq
where it should be noticed that ${\cal I}_+ \approx {\cal I}_1+{\cal I}_2$
when ${\cal I}_0\gg{\cal I}_1+{\cal I}_2$.
The components of the normal modes $\bld{s}_\pm$ may thus be sampled
from the respective Boltzmann distributions and
the expectation value of an observable $F\{\bld{S}_i\}$ can be evaluated as
\beq
\langle F\{\bld{S}_i\}\rangle = {!\over\Omega_T}
\int\!d^D\bld{s}_+ \int\!d^D\bld{s}_-\, F\{\bld{S}_i\}\, {\rm e}^{-E/T}\ ,
\eeq
where $E$ is the rotational energy given in Eq.\ (\ref{E})
and the canonical phase space is
$\Omega_T = [(2\pi{\cal I}_+T)(2\pi{\cal I}_-T)]^{D/2}$.

Once the normal spins $\bld{s}_\pm$ have been sampled,
the individual fragment spins can readily be constructed \cite{RDV},
\beq
\bld{S}_1={{\cal I}_1\over{\cal I}_1+{\cal I}_2}\bld{s}_++\bld{s}_-\ ,\,\,\
\bld{S}_2={{\cal I}_2\over{\cal I}_1+{\cal I}_2}\bld{s}_+-\bld{s}_-\ ,
\eeq
and the orbital angular momentum is $\bld{S}_0=-\bld{s}_+$.
Thus the conservation of angular momentum is built into the normal modes
$\bld{s}_\pm$, each of which carries no total angular momentum.
The mode sampling method has the special advantage 
that different temperatures can be employed for different modes, 
thus making it possible to control their relative presence,
as was recently exploited \cite{RDV}.

The sampling via normal modes is also efficient, robust, and fast.
Importantly, it yields the same ensemble of spins $\{\bld{S}_i\}$
as the direct microcanonical sampling described in Sect.\ \ref{microcan}
provided that the energy $E$ in Eq.\ (\ref{micro})
is sampled from the appropriate canonical distribution
for three coupled $D$-dimensional spins,
$P(E)\sim E^{D-1}\exp(-E/T)$.

\section{Spin opening-angle distributions}

The above sampling methods are now applied
to the calculation of the distribution of
the opening angle between the fission fragment angular momenta
for various scenarios of current interest.

\subsection{Three-dimensional spins}
\label{3D}
We consider here the scenario where the three angular momenta involved
are three-dimensional, \ie\ $D=3$, 
so we may write $\bld{S}_i=(S_i^x,S_i^y,S_i^z)$.

First, to establish a convenient reference scenario,
we assume that the two fragment spins are sampled 
entirely independently from isotropic distributions,
such as Boltzmann distributions, $P(\bld{S}_i)\sim\exp(-S_i^2/2{\cal I}_iT_i)$,
where the values of the temperature parameters $T_i$ are immaterial.
The directions of the fragment spin vectors are then distributed uniformly
over $4\pi$ and it follows that the distribution of $\cos\psi$
is constant, equivalent to the opening angle itself having the distributon
$P_{12}^{\rm indep}(\psi)\sim\sin\psi$.

In reality the fragments are interacting and their spins 
are coupled to the angular momentum of their relative motion, $\bld{S}_0$.
Because the combined system is isolated, its total angular momentum remains
unchanged and so the appropriate statistical spin distribution 
has a microcanonical form.

\begin{figure}[tbh]	       
\vspace{2in}
\includegraphics{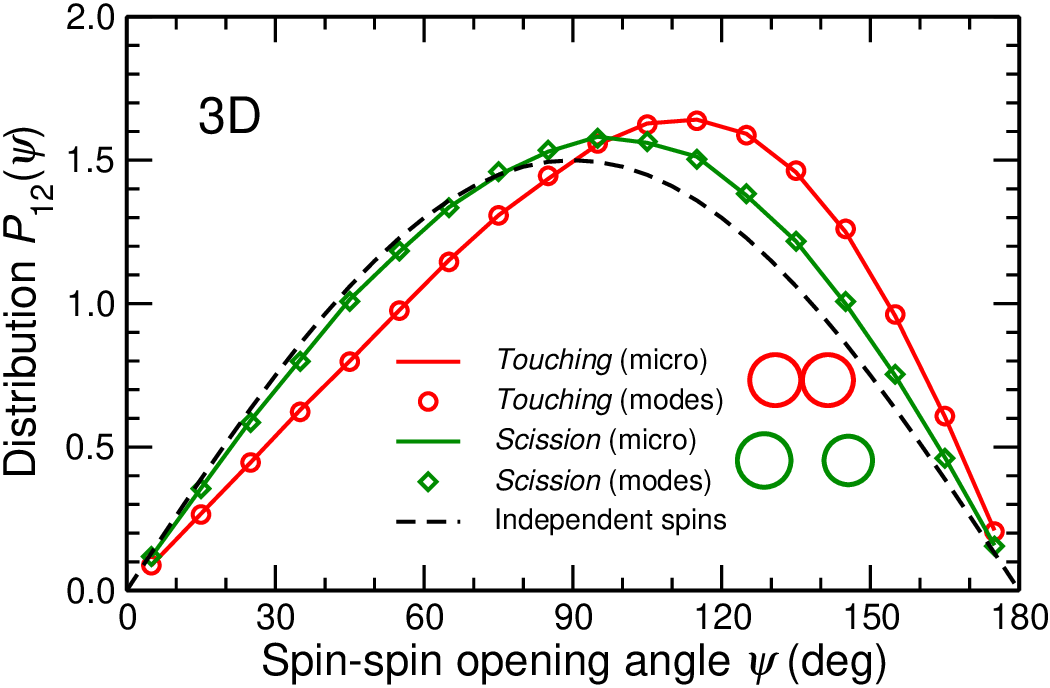}
\caption{\label{f:3D}
The distribution of the fragment spin opening angle $\psi$
obtained by 3D sampling in various scenarios:
{\em Touching:} A schematic reference scenario of
two touching spheres of equal size
in which case the relative sizes of the moments of inertia are
${\cal I}_1\!:\!{\cal I}_2\!:\!{\cal I}_0=1\!:\!1\!:\!5$;
{\em Scission:} A more realistic scenario typical of scission
for which the moments of inertia have ratios similar to those
used in \code\ (see Table \ref{t:I});
and {\em Independent:} The limiting scenario for large ${\cal I}_0$
where the angular momentum constraint is ineffective 
and the two fragment spins become independent.
Each curve is based on one million spin triplets,
obtained either by direct microcanonical sampling (Sect.\ \ref{microcan})
or by sampling of the normal modes (Sect.\ \ref{modes}).
}\end{figure}		     	

We first show results for the simple (but unrealistic) case
where the two fragments are equal in size and are touching;
the ratios of their moments of inertia are then
${\cal I}_1\!:\!{\cal I}_2\!:\!{\cal I}_0=1\!:\!1\!:\!5$.
The resulting opening-angle distribution, $P_{12}^{\rm touch}(\psi)$,
is displayed in Fig.\ \ref{f:3D}.
The coupling causes it to be skewed away from symmetry 
and it peaks near $\psi=110^\circ$.

The touching-sphere scenario is not realistic because 
the fragment formation occurs for elongated scission configurations
for which the distance between the fragment centers 
considerably exceeds the sum of the two fragment radii.
(In model calculations, the center separation 
typically exceeds the sum of the fragment radii by $d=4$~fm.)
As a consequence, the moment of inertia for the relative motion at scission
exceeds those of the individual fragments by over an order of magnitude
(whereas the individual moments of inertia are of comparable size).
In order to approximate such a scenario for illustrative purposes,
we employ moments of inertia that are similar to those used in \code,
see Table \ref{t:I},
but the present results are not very sensitive to the precise values.
[We note that $^1$) the moments of inertia employed in \code\ lead to
a reasonable reproduction of the overall fragment spin distribution
\cite{RandrupPRL127} and
$^2$) only the {\em relative} sizes of the moments of inertia are needed
for the present study,]

\begin{table}[bth]		 
\begin{tabular}{c|ccc}\hline\hline\\[-2ex]
Case &~ ${\cal I}_1/\bar{\cal I}$~ &~ ${\cal I}_2/\bar{\cal I}$~ 
     &~ ${\cal I}_0/\bar{\cal I}$~\\[1ex]
\hline\\[-2.5ex]
$^{235}$U($n_{\rm th}$,f)  &~ 0.71 &~ 1.29 &~ 16.91\\[0.5ex]
~~$^{239}$Pu($n_{\rm th}$,f)~~ &~ 0.73 &~ 1.27 &~ 17.02\\[0.5ex]
$^{252}$Cf(sf)             &~ 0.77 &~ 1.23 &~ 17.08\\[0.5ex]
\hline
This work                  &~ 0.75 &~ 1.25 &~ 17\\[0.5ex]
\hline\hline
\end{tabular}
\caption{The average values of the moments of inertia
used by the fission event generator \code\ \cite{RandrupPRC80},
relative to the mean fragment moment of inertia
$\bar{\cal I}=({\cal I}_1+{\cal I}_2)/2$.
The last line shows the ratios used here to illustrate scission.
}\label{t:I}
\end{table}		     	

The resulting opening-angle distribution, 
$P_{12}^{\rm sciss}(\psi)$, is much closer to the
limiting uncorrelated form than touching spheres, reflecting the fact that
the coupling to the relative motion becomes less effective
as the associated moment of inertia ${\cal I}_0$ is increased.

\begin{figure}[tbh]	        
\vspace{2in}
\includegraphics{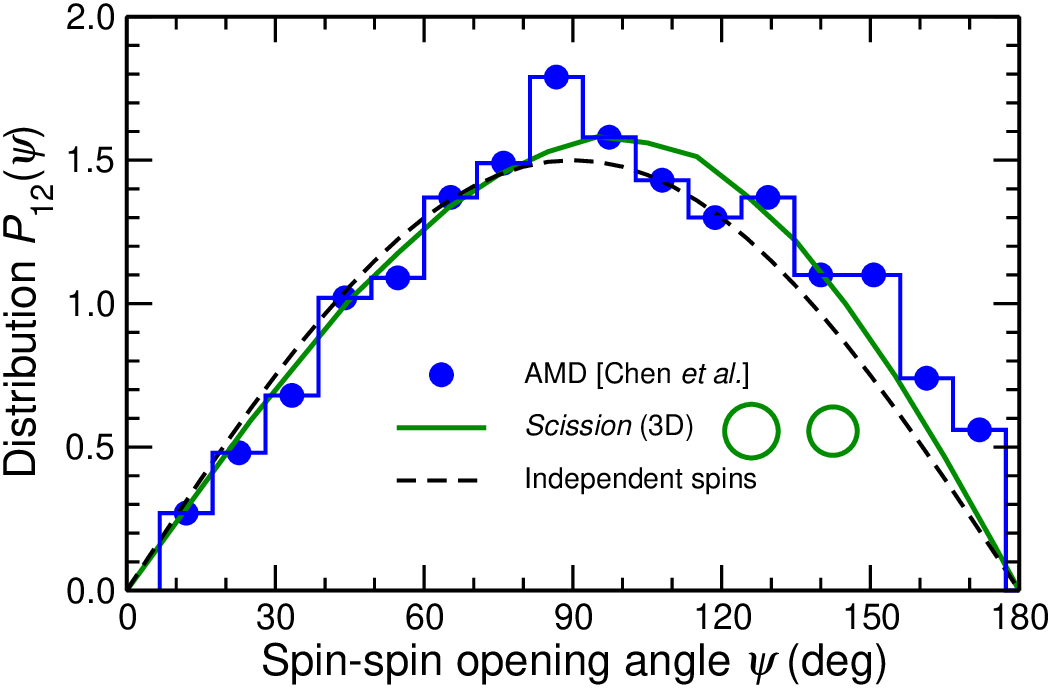}
\caption{\label{f:A}
Thw opening-angle distribution obtained by Chen, Ishizuka, and Chiba \cite{AMD}
with Anti-symmetrized Molecular Dynamics for fission of $^{252}$Cf
are compared with the 3D sampling results for the {\em scission} scenario,
$P_{12}^{\rm sciss}(\psi)$, shown in Fig.\ \ref{f:3D}.
}\end{figure}		     	

It is interesting that this distribution
quite closely resembles the one obtained by recent AMD simulations 
of fission of $^{252}$Cf \cite{AMD}, as shown in Fig.\ \ref{f:A}.
Antisymmetrized Molecular Dynamics represents the state of the system
by a Slater determinant of Gaussian wave packets 
whose centroids are propagated by classical equations of motion
with the potential energy having been augmented by 
the repulsive effect of the antisymmetrization.
This implies a short mean free path for the individual centroids 
and a rapid local equilibration might therefore be expected.
This appears to be indeed borne out by the AMD results 
for the distribution of the spin-spin opening angle 
which are consistent with the 3D equilibrium form in Fig.\ \ref{f:3D}.

\begin{figure}[tbh]	        
\vspace{2in}
\includegraphics{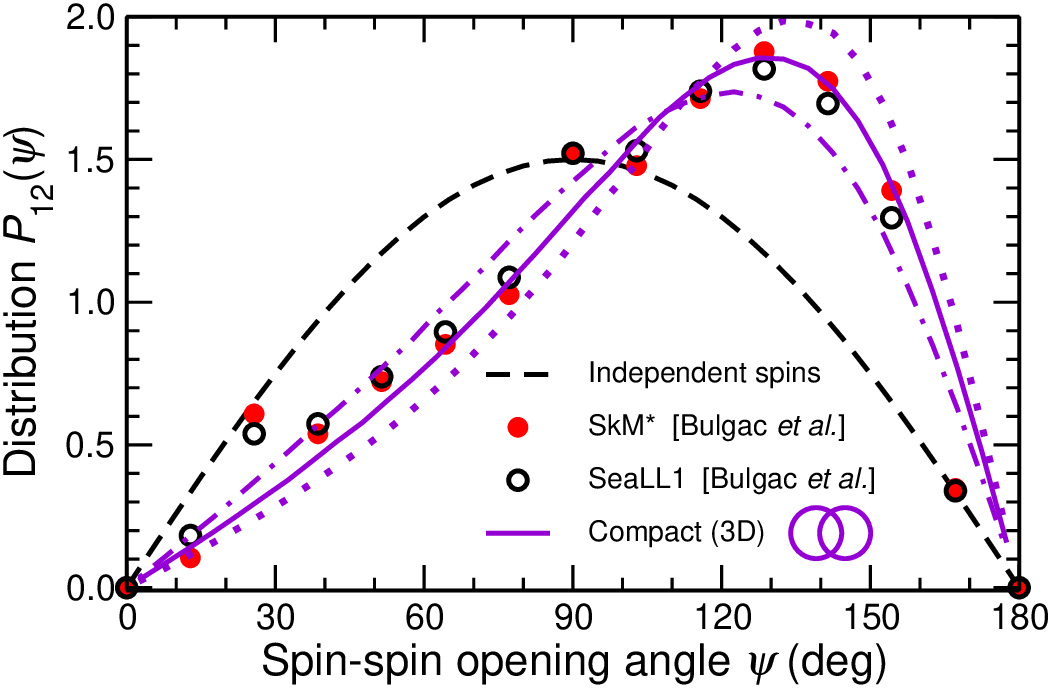}
\caption{\label{f:B}
The opening-angle distributions calculated by Bulgac \etal\ \cite{BulgacPRL128}
with time-dependent density functional theory
using either the SkM* or the SeaLL1 energy density functional
are compared with 3D sampling results for a
{\em compact} scenario for which ${\cal I}_1:{\cal I}_2:{\cal I}_0=1:1:2$.
To illustrate the sensitivity to ${\cal I}_0$ are also shown the distributions
for a 50\% smaller (dots) or a 50\% larger (dot-dash) ${\cal I}_0$ value.
}\end{figure}		     	

It is also noteworthy that the results obtained for $P_{12}(\psi)$
by Bulgac \etal\ \cite{BulgacPRL128}
using time-dependent density functional theory
can be well reproduced by 3D samplings that
employ the ratios ${\cal I}_1\!:\!{\cal I}_2\!:\!{\cal I}_0=1\!:\!1\!:\!2$,
as shown in Fig.\ \ref{f:B}.
It is quite remarkable that such a good agreement can be obtained
by using a moment of inertia for the relative motion
that is just the sum of the two individual moments of inertia,
a value that is only 40\%\ of that for touching spheres 
(about an order of magnitude below those for typical scission configurations)
and corresponds to the two fragments overlapping significantly.

To provide a sense of how well determined the relative value of ${\cal I}_0$ is,
Fig.\ \ref{f:B} also shows the distributions obtained with ${\cal I}_0$ values
that are either 50\%\ smaller (\ie\ 1:1:1) or 50\%\ larger (\ie\ 1:1:3).
Neither one of those distributions comes close to reproducing
the results from Ref.\ \cite{BulgacPRL128}.
Thus it appears that the optimal value is 
rather narrowly determined to be ${\cal I}_0\approx{\cal I}_1+{\cal I}_2$.

\subsection{Two-dimensional spins}
\label{2D}

It is theoretically expected \cite{DossingNPA433,RDV},
as well as experimentally indicated \cite{WilhelmyPRC5,WolfPRC13},
that the fission fragment angular momenta
are predominantly perpendicular to the fission axis.
It is therefore of interest to also analyze idealized scenarios
where the fragment spins are perfectly perpendicular to the fission axis.
Such a situation is analogous to the above case (Sect.\ \ref{3D}),
except that the dimensionality is now only $D=2$, 
so $\bld{S}_i=(S_i^x,S_i^y,0)$.
Figure \ref{f:2D} shows the spin-spin opening angle distribution
for the same instructive 2D scenarios as shown in Fig.\ \ref{f:3D}.

\begin{figure}[tbh]		
\label{f:2D}
\vspace{2in}
\includegraphics{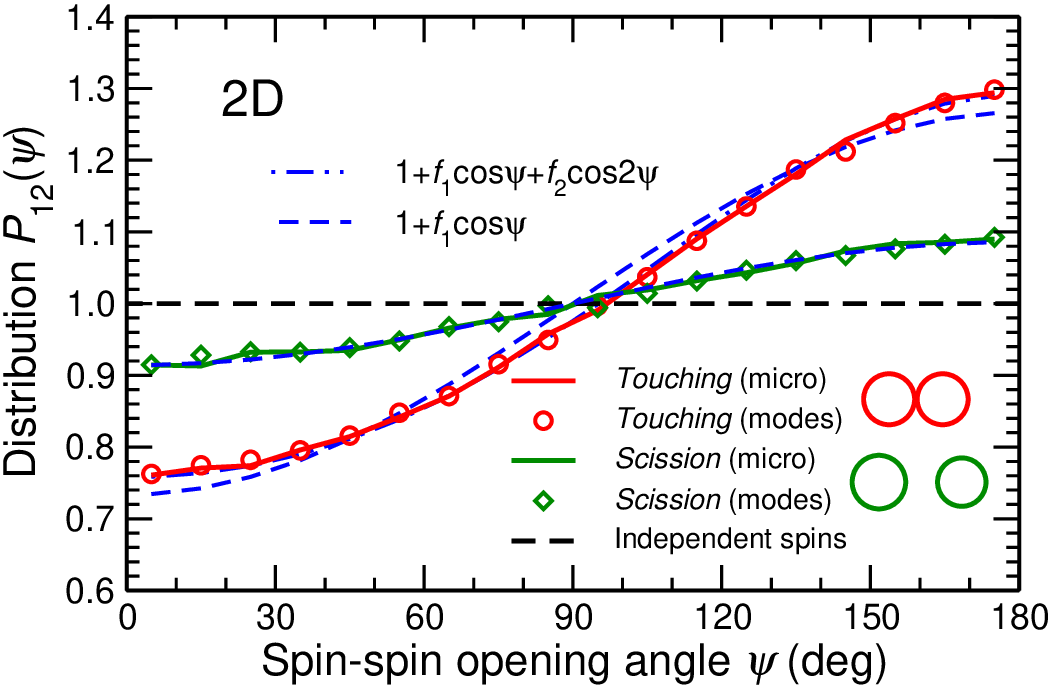}
\caption{\label{f:2D}
The distribution of the fragment spin opening angle $\psi$
obtained by 2D sampling of the three angular momenta,
using moments of inertia corresponding 
either to {\em touching} (${\cal I}_1:{\cal I}_2:{\cal I}_0=1:1:5$)
or to {\em scission} (${\cal I}_1:{\cal I}_2:{\cal I}_0=0.75\!:\!1.25\!:\!17$).
The samplings were done either microcanonically (Sect.\ \ref{microcan})
or via the normal modes (Sect.\ \ref{modes}).
The two dashed curves are the corresponding first-order Fourier fits,
while the dot-dashed curve is the second-order Fourier fit to the
touching-sphere distribution which has a larger amplitude.
}\end{figure}		     	

In the reference scenario of totally independent spins,
${\cal I}_0/\bar{\cal I}\to\infty$,
the directions of the fragment spin vectors are distributed uniformly
in the perpendicular plane
and it follows that the distribution of the opening angle $\psi$ is constant.

When the coupling to the orbital motion is taken into account in the sampling,
the two fragment spins have a slight preference for being directed oppositely,
The opening-angle distribution is typically well represented
by the lowest-order Fourier approximation,
$P_{12}(\psi)\sim1+f_1\cos\psi$.
When scission moments of inertia are used the deviation
from uniformity is fairly small, $f_1=-0.086$.

As was the case in 3D,
the touching-sphere configuration, with its considerably smaller ${\cal I}_0$,
leads to larger deviations of $P_{12}(\psi)$ from the independent scenario,
namely $f_1=-0.264$, and so the second-order Fourier term
is required for an accurate representation,
$P_{12}^{\rm touch}(\psi)\sim1+f_1\cos\psi+f_2\cos2\psi$, with $f_2=0.028$,
as is apparent from Fig.\ \ref{f:2D}.

The standard version of the fission model \code\ \cite{RandrupPRC89} 
assumes that the fission fragments emerge with angular momenta
that are perpendicular to the fission axis
and they are therefore sampled from the corresponding 2D distribution.
The resulting spin-spin opening-angle distribution \cite{VogtPRC103}
is then in accordance with the results sampled here,
as shown in Fig.\ \ref{f:F}.

\begin{figure}[tbh]		
\label{f:F}
\vspace{2in}
\includegraphics{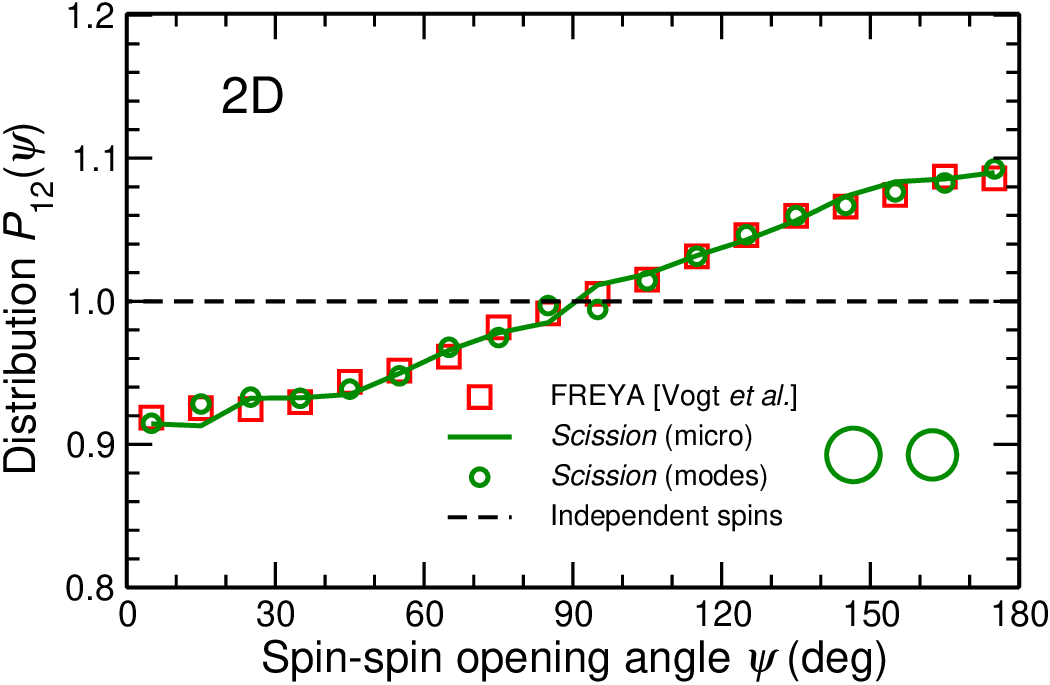}
\caption{\label{f:F}
The distribution of the fragment spin opening angle $\psi$
obtained with \code\ for $^{235}$U($n$,f)
\cite{VogtPRC103} is compared with the 2D 
sampling results for the {\em scission} scenario shown in Fig.\ \ref{f:2D}.
}\end{figure}		     	

\section{Concluding remarks}

This study describes two different but equivalent methods 
for sampling angular momenta that are correlated due to conservation laws.
These methods were applied to sampling the angular momenta of fission fragments
in either three or two dimensions.
With a focus on the distribution of the spin-spin opening angle $\psi$,
it was illustrated how the magnitude of the moment of inertia for the
relative motion influences $P_{12}(\psi)$ significantly.

Comparisons with recent model calculations of the opening-angle distribution
showed that the result obtained with Antisymmetrized Molecular Dynamics
\cite{AMD} agrees well with the statistical form pertaining to 
3D spins with moments of inertia typiocal of scission, as might be expected.
On the other hand, it is puzzling that results obtained with microscopic 
time-dependent functional theory \cite{BulgacPRL128} can be reproduced
with the 3D equlibrium distribution using a relative moment of inertia that
corresponds to a shape that is significantly more compact than touching spheres.
It may be noted that the 3D samplings do not invoke the scission geometry
and thus ignores the basic geometric requirement 
that the relative angular momentum be perpendicular to the fission axis.  

Finally, it was shown that the 2D equilibrium form
with scission moments of inertia reproduces the results of fission simulations
with the \code\ code \cite{VogtPRC103,RandrupPRL127}
which does take account of the specific scission geometry and generates
fragment spins that are perpendicular to the fission axis.

The present analysis brings out an important feature of the
coupled angular momenta appearing in fission:
The relative motion,
due to the large size of the associated moment of inertia
in comparison with those of the individual fragments,
effectively acts as a reservoir of angular momentum.
Then the conservation of angular momentum has little effect
on the fragment spins and they become nearly independent.
Indeed, the angular momenta generated by \code\ are only slightly correlated
with regard to both their directions and their magnitudes.
The latter feature was recently observed experimentally \cite{Nature}.

In view of the large differences between the model calculations
of the spin-spin opening angle distribution,
experimental information on this observable is highly desirable
as it could help to clarify the scission physics.

\section*{Acknowledgments}
We wish to acknowledge helpful communications with
T.\ D{\o}ssing, L.\ Sobotka, R.\ Vogt, and J.\ Wilson.
This work was supported by the Office of Nuclear Physics
in the U.S.\ Department of Energy's Office of Science 
under Contract No.\ DE-AC02-05CH11231;
it was stimulated by the 
Worshop on Fission Fragment Angular Momenta
recently hosted by A.\ Bulgac at the University of Washington in Seattle,
June 21-24, 2022.



                        \end{document}